\documentclass[aip,reprint,floatfix,superscriptaddress,showpacs,amsfonts
,amssymb,amsmath,preprintnumbers]{revtex4-1}
\usepackage{bm}
\usepackage{epsfig}
\usepackage{amsmath}
\usepackage{amssymb}
\usepackage{graphicx}
\usepackage{bm}
\usepackage{color}
\usepackage{subfigure}

\newcommand{\fig}[1]{Fig.~\ref{#1}}

\newcommand{\eq}[1]{Eq.~(\ref{#1})}

\newcommand{\be}{\begin{equation}}
\newcommand{\ee}{\end{equation}}

\newcommand\bea{\begin{eqnarray}}
\newcommand\eea{\end{eqnarray}}

\newcommand{\curl}{\bm \nabla \times}

\begin{document}
\setcounter{secnumdepth}{1}

\title{The generation of magnetic fields by the Biermann battery and the interplay with the Weibel instability.}

\author{K. M. Schoeffler}
\affiliation{Instituto de Plasmas e Fus\~ao Nuclear,
Instituto Superior T\'ecnico,\\ Universidade de Lisboa, 1049-001 Lisboa, Portugal}
\author{N. F. Loureiro}
\affiliation{Instituto de Plasmas e Fus\~ao Nuclear,
Instituto Superior T\'ecnico,\\ Universidade de Lisboa, 1049-001 Lisboa, Portugal}
\affiliation{Plasma Science and Fusion Center, Massachusetts Institute of Technology, Cambridge MA 02139, USA}

\author{R. A. Fonseca}
\affiliation{Instituto de Plasmas e Fus\~ao Nuclear,
Instituto Superior T\'ecnico,\\ Universidade de Lisboa, 1049-001 Lisboa, Portugal}
\affiliation{DCTI/ISCTE---Instituto Universit\'ario de Lisboa, 1649-026 Lisboa, Portugal}

\author{L. O. Silva}
\affiliation{Instituto de Plasmas e Fus\~ao Nuclear,
Instituto Superior T\'ecnico,\\ Universidade de Lisboa, 1049-001 Lisboa, Portugal}

\date{\today}

\begin{abstract}

An investigation of magnetic fields generated in an expanding bubble of plasma
with misaligned temperature and density gradients (driving the Biermann battery
mechanism) is performed. With gradient scales $L$, large-scale magnetic fields
are generated by the Biermann battery mechanism with plasma $\beta \sim 1$, as
long as $L$ is comparable to the ion inertial length $d_i$. For larger system
sizes, $L/d_e > 100$ (where $d_e$ is the electron inertial length), the Weibel
instability generates magnetic fields of similar magnitude but with wavenumber
$k d_e \approx 0.2$. In both cases, the growth and saturation of these fields have
a weak dependence on mass ratio $m_i/m_e$, indicating electron mediated physics.
A scan in system size is performed at $m_i/m_e = 2000$, showing agreement with
previous results with $m_i/m_e = 25$. In addition, the instability found at
large system sizes is quantitatively demonstrated to be the Weibel instability.
Furthermore, magnetic and electric energy spectra at scales below the electron
Larmor radius are found to exhibit power law behavior with spectral indices
$-16/3$ and $-4/3$, respectively.

\end{abstract}


\maketitle

%***********************************************************************************

\section{Introduction}
The origin of magnetic fields starting from unmagnetized plasmas is a central
question in astrophysics. Although much of the observable universe is magnetized
such that the magnetic field plays an important role in the dynamics, in the
early universe this was not so.  During the period before recombination, when
the cosmic microwave background was generated, it is widely accepted that there
was no magnetic field~\cite{Kulsrud08}. Magnetic field growth is generally
attributed to the turbulent dynamo~\cite{Kulsrud92,Brandenburg12}, which
greatly amplifies a required initial seed field. The Biermann battery mechanism
\cite{Biermann50}, in contrast, generates magnetic fields in the absence of a
seed via perpendicular density and temperature gradients, and is thought to be
the major source of these seed fields.  The Biermann mechanism can explain the
generation of a $\sim 10^{-20} G$ field after recombination, but it is
questionable whether turbulent dynamo growth alone can explain amplification up
to the $10^{-6} G$ fields seen today throughout the interstellar
medium~\cite{Kulsrud08}, (where the magnetic pressure is of the order of the
plasma pressure, $\beta\equiv 8\pi P/B^2 \sim 1$). A possible solution to this
problem may be provided by the potential role played by kinetic instabilities
in the amplification of magnetic fields.  One such instability is the Weibel
instability~\cite{Weibel59}, which we will discuss later.

The Biermann mechanism is also the presumed cause of self-generated magnetic fields (of
order $10^{6} G$, $\beta \sim 1$) found in laser-solid interaction experiments
~\cite{Stamper71,Li06,Gao15}. The laser generates an expanding bubble of
plasma by hitting and ionizing a solid foil of metal or plastic. This bubble
thus has a temperature gradient perpendicular to the beam (hottest closest to
the beam axis), and a density gradient in the direction normal to the foil,
allowing for the Biermann battery to take place. These experiments, in addition
to being interesting in themselves and in inertial confinement fusion, provide
an opportunity to help clarify poorly understood astrophysical processes,
namely magnetic field generation and amplification, and even turbulence at both
fluid and kinetic scales. Temperature gradients form perpendicular to
astrophysical shocks (hotter in the center of the shock), while density
gradients form parallel to the shock, once again allowing the Biermann battery
to take place.

Inspired by the laser-plasma configuration, we investigated, in a previous
paper~\cite{Schoeffler14}, the generation and amplification of magnetic fields
using particle-in-cell (PIC) simulations of an expanding plasma bubble with
perpendicular temperature and density gradients.  These kinetic simulations
confirmed the fluid prediction of the Biermann
battery~\cite{Max78,Craxton78,Haines97} with the generation and saturation of
magnetic fields scaling as $1/L$ for moderate values of $L/d_i>1$, where $d_i$
is the ion inertial length, and $L$ is the system size. In addition, we found
that, for large $L$ these simulations revealed the Weibel instability, which is
kinetic in nature, as the major source of magnetic fields, allowing the
magnetic field to remain finite ($\beta \sim 1$) for large $L$.  We will
hereupon refer to Ref.~\cite{Schoeffler14} as Paper I.

In the present paper a more in-depth study of this problem is performed. In the
following sections we further detail and expand the results of Paper I~\cite{Schoeffler14}, showing
the scaling of Biermann generated magnetic fields with system size, and the
formation of the Weibel instability for large system sizes ($L/d_e > 100$).  In
the next two sections we show the theory behind the two mechanisms for magnetic
field growth; the Biermann battery mechanism (section II) and the Weibel
instability (section III).  Section IV describes the computational setup. In
section V we provide evidence that the results from Paper I~\cite{Schoeffler14} hold for larger,
more realistic, mass ratios.  In Section VI we show further agreement between
observations of the instability in the large $L/d_e$ regime and theoretical
predictions of the Weibel instability.  In Section VII we present agreement with
gyrokinetic theory via power law slopes of both the magnetic and electric
field energy.  Finally, in section VIII we reiterate the importance of the $1/L$
scaling of Biermann magnetic fields, and the existence of the finite Weibel
magnetic fields ($\beta_e \sim 1$) generated in large scale temperature and
density gradients, relevant for both astrophysical systems and some current and
future laser setups.

\section{Biermann battery}

Assuming a two-fluid description of a plasma with massless electrons, the
magnetic field evolution is given by the generalized induction equation,
\begin{multline}
\label{induction}
\frac{\partial \bm{B}}{\partial t} = 
\bm \nabla \times \left(\bm v \times \bm B\right)
+ \frac{\eta c^2}{4 \pi} \nabla^2 \bm B
- \frac{1}{e n} \curl \left(\bm j \times \bm B\right)\\
- \frac{c}{n e} \bm \nabla n \times \bm \nabla T_e,
\end{multline}
which shows the evolution of magnetic field $\bm B$, based on the fluid
velocity $\bm v$, the current density $\bm j = c \bm \nabla \times \bm B / 4
\pi$, the number density $n$, and the electron temperature $T_e\equiv P_e/n$,
where $P_e$ is the electron plasma pressure.  Here, $c$ is the speed of light,
$\eta$ is the resistivity, and $e$ is the charge of an electron. 
The terms on the RHS of \eq{induction} from
left to right are the convective term, the resistive term, the Hall term, and
the Biermann battery term. The induction equation is often simplified by
assuming the system size $L$ is large compared to all kinetic scales, and only
considering the convective term on the RHS. The resistive term compared to the
convective term scales as $\delta_R/L$ where $\delta_R \equiv \eta c^2/4 \pi v$
is the resistive scale, the Hall term as $(d_i/L) (v_A/v)$ where $v_A$ is the
Alfv\'{e}n velocity, and the Biermann battery term as $(\rho_e/L) (v_{the}/v)$
where $\rho_e$ is the electron Larmor radius, and $v_{the}$ is the electron
thermal velocity. In ideal magnetohydrodynamics (MHD), all of these terms are
neglected because $\delta_R$,  $d_i$, and $\rho_e$ are assumed small compared
to $L$.  

The Biermann battery term operates when the density and the temperature
gradients are not parallel to each other ($\bm \nabla n \times \bm \nabla T_e
\ne 0$).  Although in MHD it is a small term, it is the only term independent
of $\bm B$ and thus dominates for small magnetic fields (where $\rho_e > L$).

Starting with $\bm B = 0$, all terms on the RHS except
for the Biermann term can be ignored, and thus $\bm B$ grows linearly. Based on scaling, given $T_e = m_e
v_{the}^2$, we find:
\begin{equation}
\label{Bevolution}
B\left(t\right) \approx \frac{m_ec}{e}\frac{v_{the}^2}{L_TL_n}t,
\end{equation}
where $L$ is more precisely defined by the length of the gradients ($L_n \equiv
n/\nabla n, L_T \equiv T_e/\nabla T_e$, which we will take to be comparable).
Assuming the other RHS terms remain small, the linear growth should continue
until the Biermann term disappears: within an electron transit time
$L_T/v_{the}$, hot electrons flow down the temperature gradient, while cold
electrons flow up, smoothing out the gradient, and effectively removing the
Biermann term. (Note, the density gradient, on the other hand smooths out at
the much slower sound transit time $L_n/c_s$, where $c_s \equiv \sqrt{T_e/m_i}$
is the sound speed.) Because the gradient changes with time, the magnetic field
growth eventually ceases to be linear and one must include terms proportional
to $-t^2$. However, we know that by $t= L_T/v_{the}$, without the Biermann
term, the magnetic field should saturate.  We can thus estimate the final
magnetic field as \eq{Bevolution} at $t= L_T/v_{the}$, and arrive at the condition
\begin{equation}
\label{saturation}
\rho_e \approx L_n,
\end{equation}
where $\rho_e \equiv v_{the}/\Omega_{ce}$ is the electron Larmor radius, and
$\Omega_{ce}\equiv eB/m_ec$ is the electron cyclotron frequency.  This can be rewritten in
terms of the electron plasma beta, leading to the following scaling of
the final magnetic field:
\begin{equation}
\label{bescaling}
\frac{B}{\sqrt{8\pi P_e}} = \beta_e^{-1/2}\approx \frac{1}{\sqrt{2}}\frac{d_e}{L_n}, 
\end{equation}
where $d_e \equiv c/\omega_{pe}$ is the electron inertial length, and
$\omega_{pe} \equiv \sqrt{4 \pi n e^2/m_e}$ is the electron plasma frequency.
For large $L_n/d_e$, which is typical of astrophysical systems and some
experiments, this means $B$ is small ($\beta_e$ large), in other words the
magnetic field generated can be considered insignificant. 

The validity of \eq{bescaling} rests on the assumption that the Biermann term
remains the dominant term in \eq{induction} as the magnetic field grows.
Let us check that this is indeed true.  The fluid velocity $\bm v$ should
quickly reach the sound speed $c_s$. Assuming this flow, at $t=L_T/v_{the}$,
the convection term scales as a factor of $\sqrt{m_e/m_i}$ smaller than the
Biermann term and thus should remain negligible. (For simplicity, we have now
assumed all gradients are of the same order $L \sim L_n \sim L_T$.)  For small
scales ($L < d_i$) the Hall term becomes stronger than the convection term.
Because $\bm{j} = c\curl \bm{B}/4 \pi$, at $t=L/v_{the}$ the Hall term scales
as a factor of $d_e^2/L^2$ smaller than the Biermann term, and thus also should
remain small as long as $L>d_e$. 

A different situation arises if the density and temperature gradients are
fixed, as may occur in a system with a long pulsed laser, i.e. one that
continually supplies the Biermann term, or in astrophysical shocks which can
also keep the gradients steady. In such cases, the convection term will eventually
become significant. If we assume that there are no dynamo effects which cause
the magnetic field to continue growing such that the convection term surpasses
the Biermann term, these two terms may eventually balance as magnetic fields
convect away from the Biermann source faster than the field can grow. This
balance results in the following scaling for the saturated magnetic field strength: 
\begin{equation}
\label{bscaling}
\frac{B}{\sqrt{8\pi P_e}} = \beta_e^{-1/2}\approx \frac{1}{\sqrt{2}}\frac{d_i}{L}, 
\end{equation}
where $d_i \equiv c/\omega_{pi}$ is the ion inertial length, and $\omega_{pi}
\equiv \sqrt{4 \pi n e^2/m_i}$ is the ion plasma frequency, or equivalently
$\rho_s \approx L$, where $\rho_s \equiv c_s/\Omega_{ce}$.  As in
\eq{bescaling}, the final magnetic field scales as $1/L$. 

This $d_i/L$ scaling was predicted in previous works \cite{Max78,Craxton78},
including Haines \cite{Haines97}, who also predicted that the saturated field
would peak at $L/d_i\approx 1$ and vanish for small $L/d_i$. For
these small, sub-$d_i$, scales the current is limited by microinstabilities
such as the lower hybrid drift instability and the ion acoustic instability,
which lead to a reduced magnetic field strength. This was confirmed for the
first time in Paper I~\cite{Schoeffler14}.

\begin{figure}
  \noindent\includegraphics[width=3.0in]{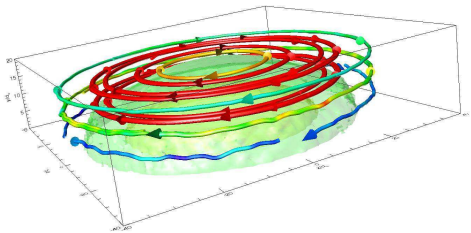}
  \caption{\label{3DBiermann}
  A still frame (at $t\omega_{pe}=235.2$, when the magnetic fields saturate) from a movie
  (multimedia view), showing the evolution of three contours of density, and a selection of magnetic field
  lines generated by the Biermann mechanism (from $t\omega_{pe}=0$ to $298.2$). This data is taken from the 3D simulation
  presented in Paper I~\cite{Schoeffler14} in which $L_T/d_e=50$ with $m_i/m_e = 25$. 
}
\end{figure}

In~\fig{3DBiermann} a still frame from a movie (multimedia view) is shown, which
illustrates this generation of magnetic fields via the Biermann battery, with
perpendicular temperature and density gradients in a 3D PIC simulation. The
profiles generating these gradients are explained in section V. Three contours
of density of the expanding plasma are displayed along with Biermann generated
magnetic field lines at a time after the magnetic fields have grown and
saturated. (This is from the 3D simulation presented in Paper
I~\cite{Schoeffler14}; here, $L/d_e=50$ and $m_i/m_e=25$.) The movie shows the
plasma expanding (at the sound speed) outward as magnetic fields are generated
via the Biermann battery mechanism. 

\section{Weibel instability}

As we showed in Paper I~\cite{Schoeffler14}, the configuration that gives rise to the Biermann
Battery also triggers the Weibel instability~\cite{Weibel59} for $L/d_e > 100$.
Note, only the temperature gradient is essential for this effect.  This
instability is caused by pressure anisotropies, and thus is an intrinsically
kinetic effect. The anisotropy in this paper is due to the kinetic evolution of
a temperature gradient in a collisionless system, the details of which will
explained in a forthcoming publication. Unlike the Biermann
battery, the Weibel instability must start from a seed magnetic field. This
seed field may be generated by the Biermann battery, or from small scale
fluctuations found in both PIC simulations and in nature, due to finite numbers
of particles (referred to as shot noise in electronics~\cite{Schottky18}).

To get a quantitative understanding of what is expected from such an
instability, let us consider an initial plasma with a bi-Maxwellian velocity
distribution:
\begin{equation}
\label{bimaxwellian}
f_{0\alpha} = n_\alpha v_{th\perp \alpha}^{-2}v_{th\parallel
\alpha}^{-1}\left(2\pi\right)^{-3/2}\exp \left(-\frac{v_{\perp \alpha}^2}{2v_{th\perp \alpha}^2} -
\frac{v_{\parallel \alpha}^2}{2v_{th\parallel \alpha}^2}\right),
\end{equation}
where the thermal velocity is larger in one direction deemed as parallel
$v_{th\parallel \alpha} > v_{th\perp \alpha}$, with $v_{th (\perp,
\parallel)\alpha} \equiv \sqrt{T_{(\perp, \parallel)\alpha}/m_{\alpha}}$. The
$\alpha$ index represents each of the particle species. 

The dispersion relation of an unmagnetized collisionless plasma with this
distribution can be obtained from the Vlasov equation. The solution for
electromagnetic perturbations consists of the
following expression~\cite{Weibel59}:
\begin{equation}
\label{solution}
k_\perp^2c^2 -\omega^2 - \sum_\alpha \omega_{p\alpha}^2 A_\alpha - \sum_\alpha
\omega_{p\alpha}^2 \left(A_\alpha+1\right) \xi_\alpha
Z\left(\xi_\alpha\right)=0
\end{equation}
where $k_\perp$ is the wavenumber, $\omega_{p\alpha}$ is the plasma frequency,
$A_\alpha \equiv T_{\parallel \alpha}/T_{\perp \alpha}-1$ is the temperature
anisotropy, $\xi_\alpha = \omega/\sqrt{2}k_\perp v_{th\perp\alpha}$, and $Z\left(\right)$ is
the plasma dispersion function~\cite{Fried61}. For the purposes of this paper
we assume that the ions do not play a role because of their large masses and low
temperature and thus we will drop the $\alpha$ index; however, in systems with
larger ion temperatures, they may play an important role on longer timescales,
especially if the electron anisotropy is not present due to collisions. In this
study, the instability is based on electron physics: the anisotropy is in the
electron pressure, and the instability forms at the electron inertial scale
$d_e$.  

There exists a purely imaginary solution where $\omega = i\gamma$ as long as
the normalized perturbation wavenumber $kd_e < \sqrt{A}$.  This unstable growing
mode is referred to as the Weibel instability, and is driven by the anisotropy
$A$.  Although we have described this instability in the context of a
bi-Maxwellian distribution, similar physics is present for a wide range of
velocity distributions with a larger velocity spread in the parallel direction.
For example, delta functions \cite{Fried59}, waterbag distributions
\cite{Yoon87}, and kappa distributions \cite{Zaheer07} all yield similar
solutions.  We will later use the bi-Maxwellian solution to quantitatively
check the growth rates and wavelengths found in our simulation.

\section{Computational Model}

Using the OSIRIS framework \cite{Fonseca02, Fonseca08}, we perform a set of
particle-in-cell (PIC) simulations to investigate the generation and
amplification of magnetic fields via the Biermann battery. We use the same
initialization for our simulations as in Paper I~\cite{Schoeffler14}, a simplification of the
aforementioned laser-plasma systems.  The fluid velocity, electric field, and
magnetic field are initially uniformly zero.

We start with a spheroid distribution of density, that has
a shorter length scale in one direction:
\begin{equation}
\begin{array} {l}
n = \begin{cases} (n_0-n_b)
\cos(\pi R_1/2L_T) + n_b, & \mbox{if } R_1 < L_T, \\
n_b, & \mbox{otherwise}, \end{cases} \\
\\
\mbox{where } R_1 = \sqrt{x^2+(L_T/L_n y)^2+z^2 },
\end{array}
\end{equation}
$n_0$ is the reference density, and $n_b=0.1n_0$ is a uniform background
density. As in section II, the characteristic lengths of the temperature and
density gradients are denoted by $L_T$ and $L_n$, respectively.  To represent
the newly formed plasma bubble, which is flatter in the direction of the
laser, $z$, we set $L_T/L_n = 2$ (this is a generic choice that appears to be
qualitatively consistent with experiments, e.g.~\cite{Nilson06,Li07,Kugland12};
note, however, that the specific value of $L_T/L_n$ depends on target and laser
properties and thus can vary).  Although this is the initial density, and can
in principle change with time, it does not evolve much during our simulations:
the density expands at the sound speed $c_s$ and all simulations are run with
$t \ll L_n/c_s$. 

The initial velocity distributions are Maxwellian, with a uniform ion thermal
velocity, $v_{th0i}$.  The spatial profile for the electron thermal velocity is
cylindrically symmetric along the $z$ direction where it is hottest in the
center. This is implemented in a similar manner to the density:
\begin{equation}
\label{cylinder}
\begin{array} {l}
v_{the} = \begin{cases} (v_{th0e}-v_{thbe}) \cos(\pi R_2/2L_T) + v_{thbe}, & \mbox{if } R_2 < L_T, \\
v_{thbe}, & \mbox{otherwise}, \end{cases} \\
\\
\mbox{where } R_2 = \sqrt{x^2+z^2},
\end{array}
\end{equation}
resulting in a maximum initial electron pressure, $P_{e0} = m_e n_0
v_{th0e}^2$. (Paper I~\cite{Schoeffler14} mistakenly stated $P_{e0} = m_e n_0 v_{th0e}^2/2$.)
The numerical values of these thermal velocities are: $v_{th0e}=0.2c$ and
$v_{th0i}=v_{thbe}=0.01c$. It is important to note that unlike the density
profile, the temperature profile changes significantly with time: the peak
drops by a factor of $\sim 2$, and the temperature gradient driving the
Biermann battery vanishes after an electron transit time. This means we should
expect agreement with \eq{bescaling} and not \eq{bscaling}. 

We will normalize the magnetic field for the results of this paper as
$B/\sqrt{8 \pi P_{e0}} = \beta_e^{-1/2}$, time as $t\omega_{pe}$ where
$\omega_{pe} \equiv \sqrt{4 \pi n_0 e^2/m_e}$, and distance as $x/d_e$ where
$d_e \equiv c/\omega_{pe}$. Unlike the previous definition of $\omega_{pe}$,
this uses the peak density $n_0$, so for regions where $n<n_0$, the local inverse
electron beta, inverse plasma frequency, and electron inertial length are
larger than the normalization indicates.

For simplicity the boundaries are periodic, but the box is large enough that
they do not interfere with the dynamics. The dimensions of the simulation
domain range from $-L,L$ in both $x$ and $y$ (and $z$ in 3D), where $L=
15/8 L_T$. The spatial resolution is $8$ grid points/$d_e$ or $1.6$ grid
points/$\lambda_d$, where $\lambda_d$ is the Debye length (This resolution is
the same as that used in Paper I~\cite{Schoeffler14}, which mistakenly reported 16 grid
points/$d_e$). The time resolution is $\Delta t\omega_{pe} = 0.07$. The 2D
simulations have 64 particles per grid cell (ppg), and the 3D simulation has 27
ppg.

\section{Insensitivity to mass ratio}

In order to investigate a larger range of $L_T/d_i$, the simulations presented
in Paper I~\cite{Schoeffler14} were run with a reduced mass ratio of $25$. In order to test the
dependence of the values reported there on mass ratio, we simulate with a mass
ratio of $100$, $400$, and the more realistic $2000$. (In most astrophysical
contexts, the mass ratio would be the 1836 of Hydrogen, while laser experiments
have a larger value.) 

\begin{figure}
  \noindent\includegraphics[width=3.0in]{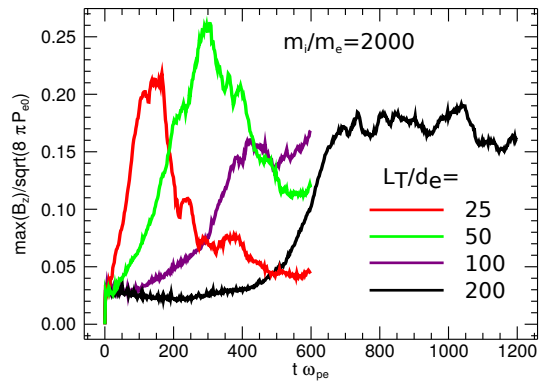}
  \caption{\label{realmassscan}
  Maximum magnetic field $B_z$ {\it vs.} time for
  a selection of system sizes ($L_T/d_e$) with $m_i/m_e=2000$. This
  figure shows the same qualitative behavior as figure 3(a) in Paper I~\cite{Schoeffler14}.
  The results of the simulation with $L_T/d_e = 400$ are presented in \fig{gammavstime}.
  }
\end{figure}

In~\fig{realmassscan}, the maximum magnetic field is displayed versus time for
various system sizes, $L_T/d_e$. This plot is equivalent to Figure 3(a) in
Paper I~\cite{Schoeffler14}, now using a realistic mass ratio of $m_i/m_e = 2000$.  There is no
qualitative difference between the realistic mass and the simulations done in
Paper I~\cite{Schoeffler14} with $m_i/m_e=25$.  For the simulations with $L_T/d_e=25$ and $50$, the
magnetic field strength reaches a peak and decays away due to the ion acoustic
instability~\cite{Haines97}. This instability is stronger at larger mass ratio, and thus still
relevant at the larger $L_T/d_e=50$. For $L_T/d_e=100$ and $200$, the magnetic
field saturates around its peak value. These are equivalent to the results
found for $m_i/m_e=25$.  The growth of the field occurs about $8$ times faster
than predicted in \eq{Bevolution}. This is an effect of our particular choice
of a density profile. The derivation of \eq{Bevolution} assumes $\bm \nabla n/n
= 1/L_n$, while more precisely  $\bm \nabla n/n = n_0/n(\bm x)L_n$, and for our
profile $n_0/n(\bm x) = 8$ at the location of fastest magnetic field growth. 

\begin{figure}
  \noindent\includegraphics[width=3.0in]{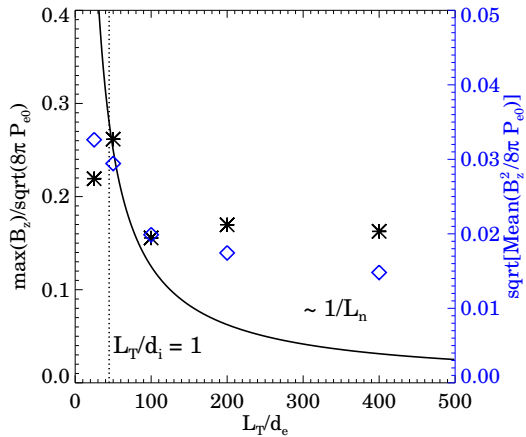}
  \caption{\label{scaling}
  Maximum (black asterisks), and average magnitude (blue diamonds) of the
  magnetic field, $B_z$, vs. $L_T/d_e$ with $m_i/m_e=2000$ at the time, $t\omega_{pe}$,
  when the magnetic field reaches a maximum.  The solid black curve indicates a
  scaling of $\sim 1/L_n$ (exactly $6.25d_e/L_n$), and the dotted line indicates where $L_T/d_i = 1$. 
}
\end{figure}
In~\fig{scaling} we present the scaling of maximum and average magnitude (the
square root of $B_z^2$ averaged in a box $2L_T\times2L_n$ surrounding the
expanding bubble) of the saturated magnetic field (or peak magnetic field for
$L_T/d_e \le 50$) as a function of system size. This is equivalent to Figure
3(b) in Paper I~\cite{Schoeffler14}, but with $m_i/m_e = 2000$. Much like the $m_i/m_e=25$
case,~\fig{scaling} shows three regions; one where the Biermann magnetic field
growth is suppressed by microinstabilities (the peak field is suppressed
for $L/d_e = 25$), a regime that scales as $d_e/L_T$ ($\approx 8d_e/\sqrt{2}L_T
\approx 6.25d_e/L_T$) after a peak around $L_T/d_i=1$, and a Weibel dominated
regime for $L/d_e \ge 100$. In contrast to the purely Biermann prediction where
the saturated magnetic field vanishes at large $L_T$, in the Weibel
dominated regime the saturated magnetic field strength remains finite for large
system sizes, confirming the conclusions obtained for $m_i/m_e=25$ in Paper I~\cite{Schoeffler14}.
The numerical trends shown in \fig{scaling} suggest that the magnetic field
amplitude at saturation in the Weibel regime is independent of $L_T$. This is
compatible with existing understanding on the saturation of the Weibel
instability~\cite{Davidson72,Califano98,Silva03}.  The presented magnitude of
the saturated magnetic field is slightly larger than shown in Paper I~\cite{Schoeffler14}
($\beta_e\approx32$ instead of $100$) (see also~\fig{depmass}). Note that at
the location of maximum  $B$, the local thermal velocity is close to
$1/2v_{th0e}$ and the local density close to $1/4n_0$. Therefore, the value of
$\beta_e$ at that location is $\beta_e\approx 2$. 

\begin{figure}
  \noindent\includegraphics[width=3.0in]{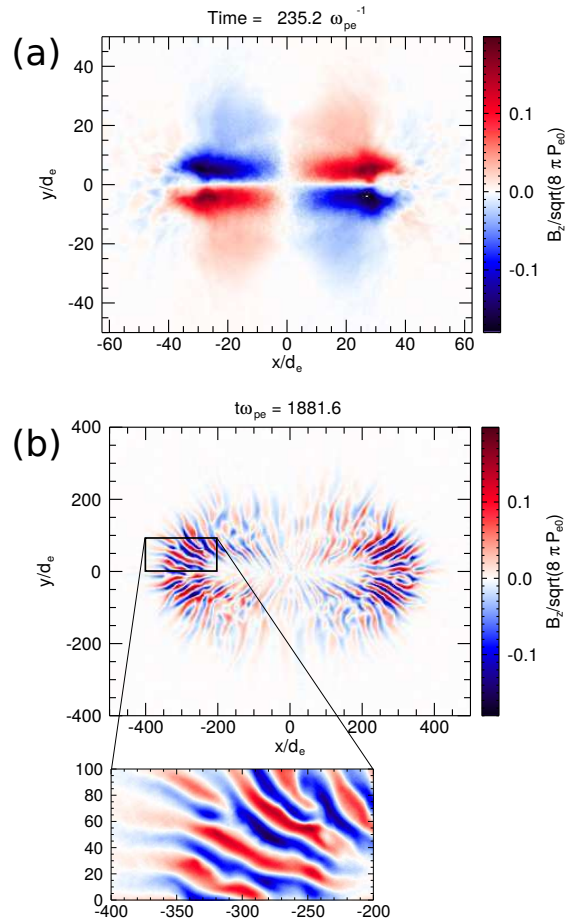}
  \caption{\label{combined2d}
    Out-of-plane magnetic field $B_z$ at peak magnitude with $m_i/m_e=2000$ for, (a) $L_T/d_e = 50$ (Biermann regime) at $t\omega_{\text{pe}} =
    235.2$, and (b) $L_T/d_e = 400$ (Weibel regime) at $t\omega_{\text{pe}} = 1881.6$, and zoom-in of Weibel fields showing $kd_e \approx 0.2$ (wavelength $\lambda/d_e \approx 31.4$).
  }
\end{figure}

The distinction between the Weibel and Biermann regimes is evident where the
saturated magnetic field strength stops following the $d_e/L_T$ scaling, but it
can be more clearly seen when observing the structure of the magnetic fields in
space for each simulation. A plot of the out-of-plane magnetic field is shown
in~\fig{combined2d}, for $L_T/d_e=50$, and $400$. In the Biermann regime, shown
in~\fig{combined2d}(a) ($L_T/d_e=50$), the spatial structures of the magnetic
field are at the system size, and follow the structures of the initial density
and temperature profiles. The Weibel regime, on the other hand, seen
in~\fig{combined2d}(b) ($L_T/d_e=400$), exhibits small ($kd_e \approx 0.2$)
magnetic structures consistent with the Weibel instability. These observations
are qualitatively the same as shown in Figure 2 of Paper I~\cite{Schoeffler14}.  The length scale
of the initial Weibel filaments is also consistent with the magnetic spectra,
which is presented (for $m_i/m_e = 25$) in Figure 5 of Paper I~\cite{Schoeffler14} showing a peak
at $kd_e \approx 0.2$.

\begin{figure}
  \noindent\includegraphics[width=3.0in]{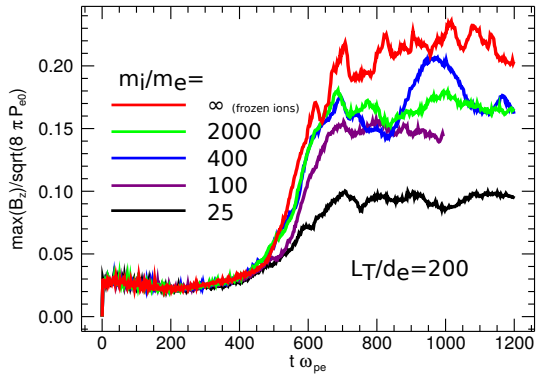}
  \caption{\label{depmass}
  Maximum magnetic field $B_z$ {\it vs.} time for
  a selection of mass ratios ($m_i/m_e$) at $L_T/d_e=200$ (Weibel regime).
  }
\end{figure}

Assuming the mass ratio is sufficiently large, the Weibel instability is
independent of mass ratio, because it is based on electron dynamics that act
at timescales much faster than the ions. We thus expect the saturated magnetic
field to asymptote as the mass ratio increases. To verify this we performed a
scan in mass ratio for the $L_T/d_e = 200$ case, which is in the Weibel regime,
but computationally small enough to perform several simulations.
In~\fig{depmass} we show that the magnitude of the magnetic field is only
weakly dependent on the mass ratio.  Although it changes in magnitude by a
factor of $2$ from $m_i/m_e =25$ to $100$, with larger mass ratios the
saturated magnetic field appears to asymptote. There is little difference even
for frozen ions where the effective mass ratio is infinite. In addition, we
find that the length scales of the magnetic perturbations are insensitive to
mass ratio.

\section{Evidence for the Weibel instability}

In Paper I~\cite{Schoeffler14} we have shown evidence of the development of the Weibel instability
in simulations of expanding plasmas with perpendicular temperature and
density gradients. Here we further show that the simulation
results exhibit excellent agreement with theoretical predictions of both the
growth rate and wavenumber for the Weibel instability. 

\begin{figure}
  \noindent\includegraphics[width=3.0in]{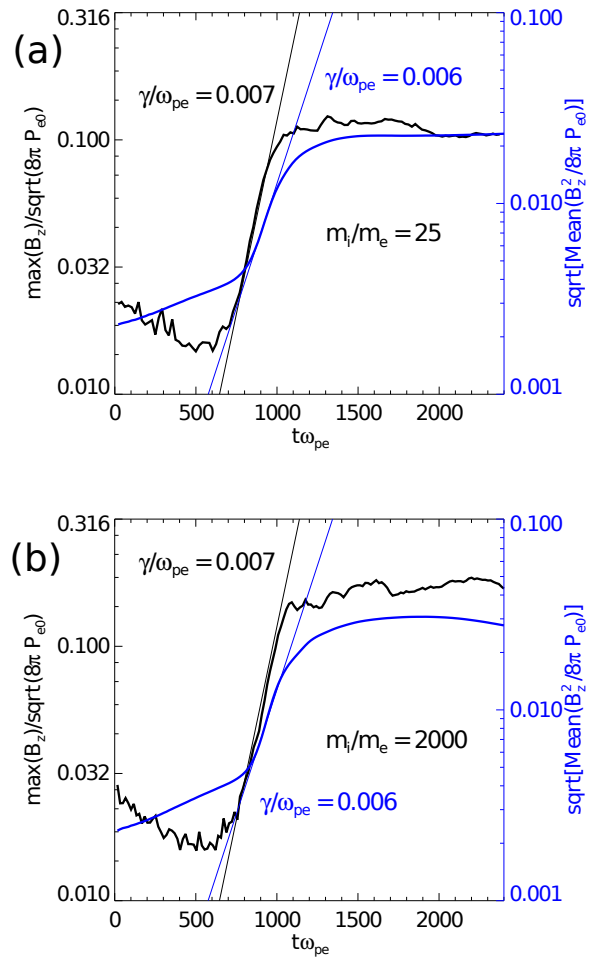}
  \caption{\label{gammameasured}
  Maximum (black curve) and average magnitude (blue curve) of magnetic field
  $B_z$ {\it vs.} time for $L_T/d_e=400$, with $m_i/m_e =$ (a) $25$, and (b)
  $2000$. This plot shows the respective growth rates (of the instability 
  identified as the Weibel instability) corresponding to
  $\gamma/\omega_{pe}=0.007$ and $0.006$.
  }
\end{figure}

Looking at the time evolution of the growth of the magnetic fields, we are able
to measure the growth rate of the instability producing the small scale
magnetic fields ($kd_e \approx 0.2$) observed in our simulations.
In~\fig{gammameasured}, for the $L_T/d_e =400$ simulation where we claim the
magnetic fields are dominated by the Weibel instability, the measured growth
rate of the maximum and average magnetic field is $\gamma/\omega_{pe} = 0.007$,
and $0.006$ respectively. We find the same measured growth rate for both the
smaller mass ratio $m_i/m_e=25$ shown in~\fig{gammameasured}(a), and the more realistic
$m_i/m_e=2000$ in~\fig{gammameasured}(b), consistent with our expectation that
the plasma dynamics we are observing are due to the electrons, not the ions.

We can test these observed growth rates and wavenumbers by comparing their
values to that predicted by theory, namely, the fastest growing mode calculated
from \eq{solution}. This is only a rough estimate of the instability because
there are density and temperature gradients in the simulation, and the velocity
distributions have evolved, and are likely not bi-Maxwellian. However, these
factors should not significantly influence the result. As stated in section
III, similar solutions exist for a variety of distribution functions, and at
the scale of the Weibel instability, the variation of temperature and density
due to the background gradients is negligible. The parameters of \eq{solution}
are: $A$, $v_{th \perp e}$, and the local plasma frequency $\omega_{pe}$, which
depends on the local plasma density $n$.  One of the outputs of OSIRIS is the
thermal spread (of the electrons) in the $x$ and $y$ directions, $v_{thex}$ and
$v_{they}$. This allows us to measure $A=v_{th \parallel e}^2/v_{th \perp
e}^2-1$ along $y=0$ where $v_{thex} = v_{th \parallel e}$, and $v_{they} =
v_{th \perp e}$. In addition, we measure $n$ so that we can solve the dispersion
relation numerically from \eq{solution} at each position along $x$. We
calculate the predicted $\gamma$ for a range of $k$, so that we can obtain the
fastest growing $\gamma$ and the associated $k$.

\begin{figure}
  \noindent\includegraphics[width=3.0in]{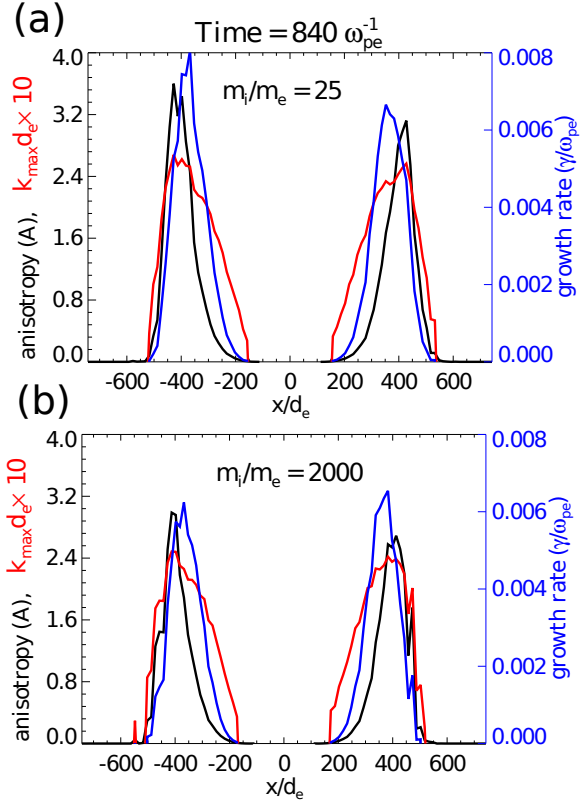}
  \caption{\label{weibeltheory}
  Anisotropy ($A$) (black curve), $k_{max}d_e$ (red curve), and growth rate
  ($\gamma/\omega_{pe}$) of the Weibel instability as predicted from \eq{solution} (blue curve), along a cut at $y=0$ at $t\omega_{pe}=840$ for
  $L_T/d_e=400$, with $m_i/m_e =$ (a) $25$, and (b) $2000$.
  }
\end{figure}
\begin{figure}
  \noindent\includegraphics[width=3.0in]{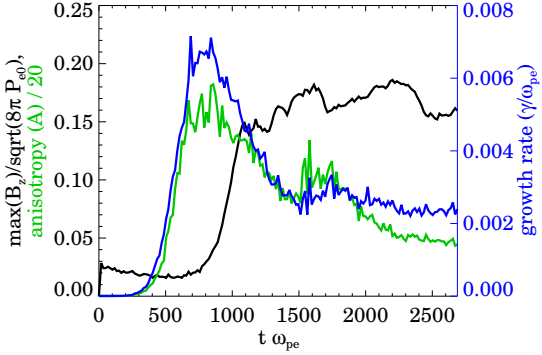}
  \caption{\label{gammavstime}
  Maximum magnetic field $B_z$ (black curve), anisotropy ($A$)/20
  (green curve), and growth rate ($\gamma/\omega_{pe}$) of the Weibel instability as predicted from \eq{solution} (blue curve), versus time for
  $L_T/d_e=400$, with $m_i/m_e = 2000$.
  }
\end{figure}
In~\fig{weibeltheory} we show a plot of the measured anisotropy $A$, and the
calculated $\gamma$ and $k$ of the fastest growing mode, along $y=0$ at
$t\omega_{pe}=840$. This is during the time with the largest growth rate
$\gamma$ ($t\omega_{pe}\approx 700-1100$, see~\fig{gammavstime}). Again we find
almost equivalent results for the smaller mass ratio
$m_i/m_e=25$ shown in ~\fig{weibeltheory}(a), and the more realistic
$m_i/m_e=2000$ in~\fig{weibeltheory}(b). The maximum growth rate along $x$ is
$\gamma/\omega_{pe} \approx 0.007$, which is consistent with the values measured above
in~\fig{gammameasured}. In addition, the calculated $kd_e \approx 0.2$ matches
the observations, and the locations with the largest growth rate
($|x/d_e|\approx 200-400$) are coincident with the Weibel fields visible
in~\fig{combined2d}(b), and Figure 2(b) of Paper I~\cite{Schoeffler14}.

In~\fig{gammavstime}, like in~\fig{realmassscan}, we show the maximum magnetic
field versus time, now for $L_T/d_e = 400$. In addition we show the maximum
measured anisotropy $A$ across $x$ (green curve), and the maximum calculated
growth rate $\gamma$ (blue curve).  The growth rate $\gamma$ increases with $A$, and
peaks at the same time as the magnetic field begins to grow, strongly
suggesting the Weibel instability. 

To summarize, there is strong evidence that the Weibel instability is generated
in our simulations; the growth rate, wavenumber, and the location and time
where the fastest growth of the measured instability occur, are consistent with
the theoretical predictions.

\begin{figure}
  \noindent\includegraphics[width=3.0in]{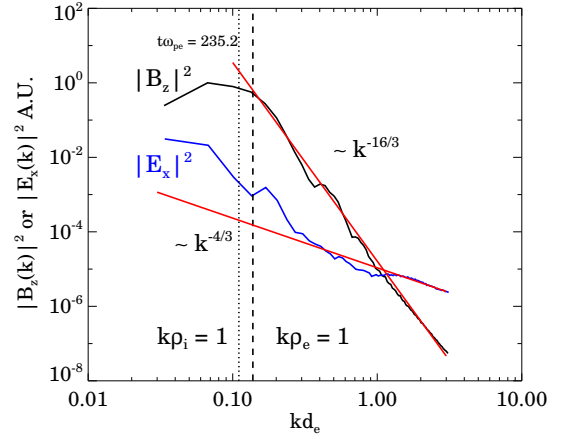}
  \caption{\label{kspect} The energy spectrum of the
  magnetic field $B_z$ (black curve), and electric field $E_x$ (blue curve),
  for the 3D simulation ($L_T/d_e=50$ and $m_i/m_e = 25$) at $t\omega_{pe}=235.2$
  (when the magnetic field growth saturates). The dotted line indicates where $k\rho_i
  =1$, and the dashed line where $k\rho_e =1$, below which power law
  behaviors are predicted, with a slope of $-16/3$ for magnetic
  energy, and $-4/3$ for electric energy, as shown in the solid red
  curves~\cite{Schekochihin09}.  [Note that in this simulation
  $\rho_i \sim \rho_e$ because the temperature ratio $T_e/T_i = 16$ is comparable
  to the mass ratio.]
}
\end{figure}

\section{Gyrokinetic power law spectra}

A surprising finding in Paper I~\cite{Schoeffler14} was the agreement of the power law slope of the
magnetic energy spectrum at scales below the electron Larmor radius, with
gyrokinetic predictions~\cite{Schekochihin09}. The prediction is based on
energy transferred via a turbulent cascade of kinetic Alfv\'{e}n waves, which
is converted at $k \rho_e \sim 1$ via electron Landau damping into an
entropy cascade.

For the 3D run that was reported in Paper I~\cite{Schoeffler14}, where $L_T/d_e=50$, the dominant
magnetic field is generated by the Biermann battery. This magnetic field (shown
in~\fig{3DBiermann}) acts as a guide field on which kinetic Alfv\'{e}n waves
may propagate, damp, and generate the predicted cascade.

In~\fig{kspect}, we show the magnetic and electric energy spectra for this run.
The spectra are obtained by performing a Fourier transform over the entire
system for one component of the electric or magnetic fields, and then averaging
over all directions of $\bm k$. In addition to the $-16/3$ power-law slope of
the magnetic energy already revealed in Paper I~\cite{Schoeffler14}, we observe that the electric
energy spectrum is in good agreement with a $-4/3$ power-law, as
predicted~\cite{Schekochihin09}. At the time shown, the spectra has reached a
steady state and remains unchanged for the rest of the simulation, up to
$t\omega_{pe}=298.2$. 

\section{Conclusions}

We have performed kinetic simulations of plasmas with perpendicular temperature
and density gradients of ranging mass ratio up to $m_i/m_e = 2000$, and
confirmed the magnetic field generation and amplification shown in Paper
I~\cite{Schoeffler14}. We have shown that as the system size increases beyond
$L_T/d_e \sim 100$, the major source of magnetic field changes from the
Biermann battery mechanism to kinetic Weibel generated fields, which remain
finite ($\beta_e \sim 1$) for large system sizes. We have done further analysis
on the simulations done in Paper I~\cite{Schoeffler14}, and on the new simulations with $m_i/m_e =
2000$ confirming that in large system sizes, magnetic fields are generated by
the Weibel instability.  Finally we have shown that our results agree with
gyrokinetic models in power laws found in both electric and magnetic energy
spectra. 

We reiterate the main applications of our conclusions shown in Paper I~\cite{Schoeffler14}, which
are significant to the topic of the generation of magnetic fields in
astrophysical contexts including shocks found in supernova remnants. Although
small scale, the Weibel instability generates magnetic fields at significant
levels ($\beta_e \sim 1$ independent of $L_T$). These are astrophysically
relevant magnitudes in typical contexts (where $L_T/d_i \gg 1$), unlike the
fields predicted by the Biermann battery that yield $\beta_e \sim
(d_i/L_T)^{-2}$.  However the discrepancy between the large scales of
astrophysical magnetic fields and the microscopic Weibel fields remains.
Whether it is possible for a mechanism such as turbulence to bridge this gap
remains an outstanding mystery of astrophysical research.

The matching of power law spectra points to the exciting possibility of validating
theories of kinetic turbulence in laser experiments. Due to the small scale separation
between $L_T$, $\rho_i$, and $\rho_e$ in our 3D simulation, we are only able to
measure a sub-$\rho_e$ power law. However, the spectra at other scales
could potentially be tested in simulations and experiments.  In addition
to the configuration that we have studied, Weibel-generated turbulence may occur due to anisotropies
induced by other mechanisms; for example, laser-generated hot electron beams
with return currents. It might be possible to probe such magnetized laser plasmas in
order to investigate turbulence at kinetic scales. 
Indeed, one such an experiment has recently been performed at the Tata Institute of Fundamental Research,
Mumbai~\cite{Mondal12}, probing the magnetic energy spectra in the
range with $\rho_i^{-1} \lesssim  k < \rho_e^{-1}$.  A power law of $-7/3$
has been measured from the experimental results (to be published shortly), consistent with theoretical
predictions~\cite{Schekochihin09}. 

In existing (and future) laser-plasma experiments the predicted $1/L$ scalings
in \eq{bescaling} and \eq{bscaling} of the saturated Biermann magnetic fields
should be relevant. Furthermore, it may be possible to find the electron-Weibel
instability as we have done here, as long as collisions do not occur faster
than the electron transit time and keep the temperature isotropic ($\nu_{ei}
L_T/v_{the} \ll 1$). Based on experimentally relevant parameters, this
condition is equivalent to the following engineering formula:
\begin{equation}
1\times10^{-2}\left(\frac{n}{1\times10^{19}\text{ cm}^{-3}}\right) \left(\frac{\ln \Lambda}{10}\right)  \left(\frac{L_T}{400 \mu \text{m}}\right) \left(\frac{T_e}{1\text{keV}}\right)^{-2} \ll 1 \text{,}
\end{equation}
where $n$ is the number density, $\ln \Lambda$ is the Coulomb logarithm, $L_T$
is the gradient scale, and $T_e$ is the electron temperature. This inequality
suggests the possibility of observing the Weibel instability, since the
physical values quoted above are reasonable numbers which can occur at the low
density, high temperature coronae generated around laser targets,
e.g.~\cite{Gao13}. Note that in this work the strong temperature gradient is
assumed to form before the anisotropy and the subsequent instability. We think
this is a good modeling assumption in cases where intense short-pulse
lasers are used. Cases with longer, less intense lasers require more careful
consideration due to the possible influence of collisions and the laser fields
during the timescale of the experiment. \smallskip 

\section{Acknowledgments}

This work was supported by the European Research Council (ERC-2010-AdG Grant
No. 267841). NFL was partially supported by the Funda\c{c}\~ao para a Ci\^{e}ncia e
Tecnologia via Grants UID/FIS/50010/2013 and IF/00530/2013.  Simulations were
carried out at SuperMUC (Germany) under a PRACE grant, and the authors
acknowledge the Gauss Centre for Supercomputing (GCS) for providing computing
time through the John von Neumann Institute for Computing (NIC) on the GCS
share of the supercomputer JUQUEEN at J\"ulich Supercomputing Centre (JSC).

%-----------------------------------------------------------------------------------

%


\begin{thebibliography}{28}%
\makeatletter
\providecommand \@ifxundefined [1]{%
 \@ifx{#1\undefined}
}%
\providecommand \@ifnum [1]{%
 \ifnum #1\expandafter \@firstoftwo
 \else \expandafter \@secondoftwo
 \fi
}%
\providecommand \@ifx [1]{%
 \ifx #1\expandafter \@firstoftwo
 \else \expandafter \@secondoftwo
 \fi
}%
\providecommand \natexlab [1]{#1}%
\providecommand \enquote  [1]{``#1''}%
\providecommand \bibnamefont  [1]{#1}%
\providecommand \bibfnamefont [1]{#1}%
\providecommand \citenamefont [1]{#1}%
\providecommand \href@noop [0]{\@secondoftwo}%
\providecommand \href [0]{\begingroup \@sanitize@url \@href}%
\providecommand \@href[1]{\@@startlink{#1}\@@href}%
\providecommand \@@href[1]{\endgroup#1\@@endlink}%
\providecommand \@sanitize@url [0]{\catcode `\\12\catcode `\$12\catcode
  `\&12\catcode `\#12\catcode `\^12\catcode `\_12\catcode `\%12\relax}%
\providecommand \@@startlink[1]{}%
\providecommand \@@endlink[0]{}%
\providecommand \url  [0]{\begingroup\@sanitize@url \@url }%
\providecommand \@url [1]{\endgroup\@href {#1}{\urlprefix }}%
\providecommand \urlprefix  [0]{URL }%
\providecommand \Eprint [0]{\href }%
\providecommand \doibase [0]{http://dx.doi.org/}%
\providecommand \selectlanguage [0]{\@gobble}%
\providecommand \bibinfo  [0]{\@secondoftwo}%
\providecommand \bibfield  [0]{\@secondoftwo}%
\providecommand \translation [1]{[#1]}%
\providecommand \BibitemOpen [0]{}%
\providecommand \bibitemStop [0]{}%
\providecommand \bibitemNoStop [0]{.\EOS\space}%
\providecommand \EOS [0]{\spacefactor3000\relax}%
\providecommand \BibitemShut  [1]{\csname bibitem#1\endcsname}%
\let\auto@bib@innerbib\@empty
%</preamble>
\bibitem [{\citenamefont {{Kulsrud}}\ and\ \citenamefont
  {{Zweibel}}(2008)}]{Kulsrud08}%
  \BibitemOpen
  \bibfield  {author} {\bibinfo {author} {\bibfnamefont {R.~M.}\ \bibnamefont
  {{Kulsrud}}}\ and\ \bibinfo {author} {\bibfnamefont {E.~G.}\ \bibnamefont
  {{Zweibel}}},\ }\href {\doibase 10.1088/0034-4885/71/4/046901} {\bibfield
  {journal} {\bibinfo  {journal} {Rep. Prog. Phys.}\ }\textbf {\bibinfo
  {volume} {71}},\ \bibinfo {eid} {046901} (\bibinfo {year}
  {2008})}\BibitemShut {NoStop}%
\bibitem [{\citenamefont {{Kulsrud}}\ and\ \citenamefont
  {{Anderson}}(1992)}]{Kulsrud92}%
  \BibitemOpen
  \bibfield  {author} {\bibinfo {author} {\bibfnamefont {R.~M.}\ \bibnamefont
  {{Kulsrud}}}\ and\ \bibinfo {author} {\bibfnamefont {S.~W.}\ \bibnamefont
  {{Anderson}}},\ }\href {\doibase 10.1086/171743} {\bibfield  {journal}
  {\bibinfo  {journal} {Astrophys. J.}\ }\textbf {\bibinfo {volume} {396}},\
  \bibinfo {pages} {606} (\bibinfo {year} {1992})}\BibitemShut {NoStop}%
\bibitem [{\citenamefont {{Brandenburg}}\ \emph {et~al.}(2012)\citenamefont
  {{Brandenburg}}, \citenamefont {{Sokoloff}},\ and\ \citenamefont
  {{Subramanian}}}]{Brandenburg12}%
  \BibitemOpen
  \bibfield  {author} {\bibinfo {author} {\bibfnamefont {A.}~\bibnamefont
  {{Brandenburg}}}, \bibinfo {author} {\bibfnamefont {D.}~\bibnamefont
  {{Sokoloff}}}, \ and\ \bibinfo {author} {\bibfnamefont {K.}~\bibnamefont
  {{Subramanian}}},\ }\href {\doibase 10.1007/s11214-012-9909-x} {\bibfield
  {journal} {\bibinfo  {journal} {Space Sci. Rev.}\ }\textbf {\bibinfo {volume}
  {169}},\ \bibinfo {pages} {123} (\bibinfo {year} {2012})}\BibitemShut
  {NoStop}%
\bibitem [{\citenamefont {Biermann}(1950)}]{Biermann50}%
  \BibitemOpen
  \bibfield  {author} {\bibinfo {author} {\bibfnamefont {L.}~\bibnamefont
  {Biermann}},\ }\href@noop {} {\bibfield  {journal} {\bibinfo  {journal} {Z.
  Naturforsch.}\ }\textbf {\bibinfo {volume} {5a}},\ \bibinfo {pages} {65}
  (\bibinfo {year} {1950})}\BibitemShut {NoStop}%
\bibitem [{\citenamefont {Weibel}(1959)}]{Weibel59}%
  \BibitemOpen
  \bibfield  {author} {\bibinfo {author} {\bibfnamefont {E.~S.}\ \bibnamefont
  {Weibel}},\ }\href {\doibase 10.1103/PhysRev.114.18} {\bibfield  {journal}
  {\bibinfo  {journal} {Phys. Rev.}\ }\textbf {\bibinfo {volume} {114}},\
  \bibinfo {pages} {18} (\bibinfo {year} {1959})}\BibitemShut {NoStop}%
\bibitem [{\citenamefont {Stamper}\ \emph {et~al.}(1971)\citenamefont
  {Stamper}, \citenamefont {Papadopoulos}, \citenamefont {Sudan}, \citenamefont
  {Dean},\ and\ \citenamefont {McLean}}]{Stamper71}%
  \BibitemOpen
  \bibfield  {author} {\bibinfo {author} {\bibfnamefont {J.~A.}\ \bibnamefont
  {Stamper}}, \bibinfo {author} {\bibfnamefont {K.}~\bibnamefont
  {Papadopoulos}}, \bibinfo {author} {\bibfnamefont {R.~N.}\ \bibnamefont
  {Sudan}}, \bibinfo {author} {\bibfnamefont {S.~O.}\ \bibnamefont {Dean}}, \
  and\ \bibinfo {author} {\bibfnamefont {E.~A.}\ \bibnamefont {McLean}},\
  }\href@noop {} {\bibfield  {journal} {\bibinfo  {journal} {Phys. Rev. Lett.}\
  }\textbf {\bibinfo {volume} {26}},\ \bibinfo {pages} {1012} (\bibinfo {year}
  {1971})}\BibitemShut {NoStop}%
\bibitem [{\citenamefont {{Li}}\ \emph {et~al.}(2006)\citenamefont {{Li}},
  \citenamefont {{S{\'e}guin}}, \citenamefont {{Frenje}}, \citenamefont
  {{Rygg}}, \citenamefont {{Petrasso}}, \citenamefont {{Town}}, \citenamefont
  {{Amendt}}, \citenamefont {{Hatchett}}, \citenamefont {{Landen}},
  \citenamefont {{MacKinnon}} \emph {et~al.}}]{Li06}%
  \BibitemOpen
  \bibfield  {author} {\bibinfo {author} {\bibfnamefont {C.~K.}\ \bibnamefont
  {{Li}}}, \bibinfo {author} {\bibfnamefont {F.~H.}\ \bibnamefont
  {{S{\'e}guin}}}, \bibinfo {author} {\bibfnamefont {J.~A.}\ \bibnamefont
  {{Frenje}}}, \bibinfo {author} {\bibfnamefont {J.~R.}\ \bibnamefont
  {{Rygg}}}, \bibinfo {author} {\bibfnamefont {R.~D.}\ \bibnamefont
  {{Petrasso}}}, \bibinfo {author} {\bibfnamefont {R.~P.~J.}\ \bibnamefont
  {{Town}}}, \bibinfo {author} {\bibfnamefont {P.~A.}\ \bibnamefont
  {{Amendt}}}, \bibinfo {author} {\bibfnamefont {S.~P.}\ \bibnamefont
  {{Hatchett}}}, \bibinfo {author} {\bibfnamefont {O.~L.}\ \bibnamefont
  {{Landen}}}, \bibinfo {author} {\bibfnamefont {A.~J.}\ \bibnamefont
  {{MacKinnon}}},  \emph {et~al.},\ }\href {\doibase
  10.1103/PhysRevLett.97.135003} {\bibfield  {journal} {\bibinfo  {journal}
  {Phys. Rev. Lett.}\ }\textbf {\bibinfo {volume} {97}},\ \bibinfo {eid}
  {135003} (\bibinfo {year} {2006})}\BibitemShut {NoStop}%
\bibitem [{\citenamefont {Gao}\ \emph {et~al.}(2015)\citenamefont {Gao},
  \citenamefont {Nilson}, \citenamefont {Igumenshchev}, \citenamefont {Haines},
  \citenamefont {Froula}, \citenamefont {Betti},\ and\ \citenamefont
  {Meyerhofer}}]{Gao15}%
  \BibitemOpen
  \bibfield  {author} {\bibinfo {author} {\bibfnamefont {L.}~\bibnamefont
  {Gao}}, \bibinfo {author} {\bibfnamefont {P.~M.}\ \bibnamefont {Nilson}},
  \bibinfo {author} {\bibfnamefont {I.~V.}\ \bibnamefont {Igumenshchev}},
  \bibinfo {author} {\bibfnamefont {M.~G.}\ \bibnamefont {Haines}}, \bibinfo
  {author} {\bibfnamefont {D.~H.}\ \bibnamefont {Froula}}, \bibinfo {author}
  {\bibfnamefont {R.}~\bibnamefont {Betti}}, \ and\ \bibinfo {author}
  {\bibfnamefont {D.~D.}\ \bibnamefont {Meyerhofer}},\ }\href {\doibase
  10.1103/PhysRevLett.114.215003} {\bibfield  {journal} {\bibinfo  {journal}
  {Phys. Rev. Lett.}\ }\textbf {\bibinfo {volume} {114}},\ \bibinfo {pages}
  {215003} (\bibinfo {year} {2015})}\BibitemShut {NoStop}%
\bibitem [{\citenamefont {Schoeffler}\ \emph {et~al.}(2014)\citenamefont
  {Schoeffler}, \citenamefont {Loureiro}, \citenamefont {Fonseca},\ and\
  \citenamefont {Silva}}]{Schoeffler14}%
  \BibitemOpen
  \bibfield  {author} {\bibinfo {author} {\bibfnamefont {K.~M.}\ \bibnamefont
  {Schoeffler}}, \bibinfo {author} {\bibfnamefont {N.~F.}\ \bibnamefont
  {Loureiro}}, \bibinfo {author} {\bibfnamefont {R.~A.}\ \bibnamefont
  {Fonseca}}, \ and\ \bibinfo {author} {\bibfnamefont {L.~O.}\ \bibnamefont
  {Silva}},\ }\href {\doibase 10.1103/PhysRevLett.112.175001} {\bibfield
  {journal} {\bibinfo  {journal} {Phys. Rev. Lett.}\ }\textbf {\bibinfo
  {volume} {112}},\ \bibinfo {pages} {175001} (\bibinfo {year}
  {2014})}\BibitemShut {NoStop}%
\bibitem [{\citenamefont {{Max}}\ \emph {et~al.}(1978)\citenamefont {{Max}},
  \citenamefont {{Thomson}},\ and\ \citenamefont {{Manheimer}}}]{Max78}%
  \BibitemOpen
  \bibfield  {author} {\bibinfo {author} {\bibfnamefont {C.~E.}\ \bibnamefont
  {{Max}}}, \bibinfo {author} {\bibfnamefont {J.~J.}\ \bibnamefont
  {{Thomson}}}, \ and\ \bibinfo {author} {\bibfnamefont {W.~M.}\ \bibnamefont
  {{Manheimer}}},\ }\href {\doibase 10.1063/1.862074} {\bibfield  {journal}
  {\bibinfo  {journal} {Phys. Fluids}\ }\textbf {\bibinfo {volume} {21}},\
  \bibinfo {pages} {128} (\bibinfo {year} {1978})}\BibitemShut {NoStop}%
\bibitem [{\citenamefont {{Craxton}}\ and\ \citenamefont
  {{Haines}}(1978)}]{Craxton78}%
  \BibitemOpen
  \bibfield  {author} {\bibinfo {author} {\bibfnamefont {R.~S.}\ \bibnamefont
  {{Craxton}}}\ and\ \bibinfo {author} {\bibfnamefont {M.~G.}\ \bibnamefont
  {{Haines}}},\ }\href {\doibase 10.1088/0032-1028/20/6/001} {\bibfield
  {journal} {\bibinfo  {journal} {Plasma Phys.}\ }\textbf {\bibinfo {volume}
  {20}},\ \bibinfo {pages} {487} (\bibinfo {year} {1978})}\BibitemShut
  {NoStop}%
\bibitem [{\citenamefont {Haines}(1997)}]{Haines97}%
  \BibitemOpen
  \bibfield  {author} {\bibinfo {author} {\bibfnamefont {M.~G.}\ \bibnamefont
  {Haines}},\ }\href@noop {} {\bibfield  {journal} {\bibinfo  {journal} {Phys.
  Rev. Lett.}\ }\textbf {\bibinfo {volume} {78}},\ \bibinfo {pages} {254}
  (\bibinfo {year} {1997})}\BibitemShut {NoStop}%
\bibitem [{\citenamefont {Schottky}(1918)}]{Schottky18}%
  \BibitemOpen
  \bibfield  {author} {\bibinfo {author} {\bibfnamefont {W.}~\bibnamefont
  {Schottky}},\ }\href {\doibase 10.1002/andp.19183622304} {\bibfield
  {journal} {\bibinfo  {journal} {Ann. Phys.}\ }\textbf {\bibinfo {volume}
  {362}},\ \bibinfo {pages} {541} (\bibinfo {year} {1918})}\BibitemShut
  {NoStop}%
\bibitem [{\citenamefont {Fried}\ and\ \citenamefont {Conte}(1961)}]{Fried61}%
  \BibitemOpen
  \bibfield  {author} {\bibinfo {author} {\bibfnamefont {B.~D.}\ \bibnamefont
  {Fried}}\ and\ \bibinfo {author} {\bibfnamefont {S.~D.}\ \bibnamefont
  {Conte}},\ }\href@noop {} {\emph {\bibinfo {title} {The Plasma Dispersion
  Function}}}\ (\bibinfo  {publisher} {Academic Press},\ \bibinfo {address}
  {New York},\ \bibinfo {year} {1961})\BibitemShut {NoStop}%
\bibitem [{\citenamefont {{Fried}}(1959)}]{Fried59}%
  \BibitemOpen
  \bibfield  {author} {\bibinfo {author} {\bibfnamefont {B.~D.}\ \bibnamefont
  {{Fried}}},\ }\href {\doibase 10.1063/1.1705933} {\bibfield  {journal}
  {\bibinfo  {journal} {Phys. Fluids}\ }\textbf {\bibinfo {volume} {2}},\
  \bibinfo {pages} {337} (\bibinfo {year} {1959})}\BibitemShut {NoStop}%
\bibitem [{\citenamefont {{Yoon}}\ and\ \citenamefont
  {{Davidson}}(1987)}]{Yoon87}%
  \BibitemOpen
  \bibfield  {author} {\bibinfo {author} {\bibfnamefont {P.~H.}\ \bibnamefont
  {{Yoon}}}\ and\ \bibinfo {author} {\bibfnamefont {R.~C.}\ \bibnamefont
  {{Davidson}}},\ }\href {\doibase 10.1103/PhysRevA.35.2718} {\bibfield
  {journal} {\bibinfo  {journal} {Phys. Rev. A}\ }\textbf {\bibinfo {volume}
  {35}},\ \bibinfo {pages} {2718} (\bibinfo {year} {1987})}\BibitemShut
  {NoStop}%
\bibitem [{\citenamefont {{Zaheer}}\ and\ \citenamefont
  {{Murtaza}}(2007)}]{Zaheer07}%
  \BibitemOpen
  \bibfield  {author} {\bibinfo {author} {\bibfnamefont {S.}~\bibnamefont
  {{Zaheer}}}\ and\ \bibinfo {author} {\bibfnamefont {G.}~\bibnamefont
  {{Murtaza}}},\ }\href {\doibase 10.1063/1.2536159} {\bibfield  {journal}
  {\bibinfo  {journal} {Phys. Plasmas}\ }\textbf {\bibinfo {volume} {14}},\
  \bibinfo {eid} {022108} (\bibinfo {year} {2007})}\BibitemShut {NoStop}%
\bibitem [{\citenamefont {Fonseca}\ \emph {et~al.}(2002)\citenamefont
  {Fonseca}, \citenamefont {Silva}, \citenamefont {Tsung}, \citenamefont
  {Decyk}, \citenamefont {Lu}, \citenamefont {Ren}, \citenamefont {Mori},
  \citenamefont {Deng}, \citenamefont {Lee}, \citenamefont {Katsouleas},\ and\
  \citenamefont {Adam}}]{Fonseca02}%
  \BibitemOpen
  \bibfield  {author} {\bibinfo {author} {\bibfnamefont {R.~A.}\ \bibnamefont
  {Fonseca}}, \bibinfo {author} {\bibfnamefont {L.~O.}\ \bibnamefont {Silva}},
  \bibinfo {author} {\bibfnamefont {F.~S.}\ \bibnamefont {Tsung}}, \bibinfo
  {author} {\bibfnamefont {V.~K.}\ \bibnamefont {Decyk}}, \bibinfo {author}
  {\bibfnamefont {W.}~\bibnamefont {Lu}}, \bibinfo {author} {\bibfnamefont
  {C.}~\bibnamefont {Ren}}, \bibinfo {author} {\bibfnamefont {W.~B.}\
  \bibnamefont {Mori}}, \bibinfo {author} {\bibfnamefont {S.}~\bibnamefont
  {Deng}}, \bibinfo {author} {\bibfnamefont {S.}~\bibnamefont {Lee}}, \bibinfo
  {author} {\bibfnamefont {T.}~\bibnamefont {Katsouleas}}, \ and\ \bibinfo
  {author} {\bibfnamefont {J.~C.}\ \bibnamefont {Adam}},\ }\href
  {http://link.springer-ny.com/link/service/series/0558/bibs/2331/23310342.htm;
  http://link.springer-ny.com/link/service/series/0558/papers/2331/23310342.pdf}
  {\bibfield  {journal} {\bibinfo  {journal} {Lect. Notes Comput. Sci.}\
  }\textbf {\bibinfo {volume} {2331}},\ \bibinfo {pages} {342} (\bibinfo {year}
  {2002})}\BibitemShut {NoStop}%
\bibitem [{\citenamefont {{Fonseca}}\ \emph {et~al.}(2008)\citenamefont
  {{Fonseca}}, \citenamefont {{Martins}}, \citenamefont {{Silva}},
  \citenamefont {{Tonge}}, \citenamefont {{Tsung}},\ and\ \citenamefont
  {{Mori}}}]{Fonseca08}%
  \BibitemOpen
  \bibfield  {author} {\bibinfo {author} {\bibfnamefont {R.~A.}\ \bibnamefont
  {{Fonseca}}}, \bibinfo {author} {\bibfnamefont {S.~F.}\ \bibnamefont
  {{Martins}}}, \bibinfo {author} {\bibfnamefont {L.~O.}\ \bibnamefont
  {{Silva}}}, \bibinfo {author} {\bibfnamefont {J.~W.}\ \bibnamefont
  {{Tonge}}}, \bibinfo {author} {\bibfnamefont {F.~S.}\ \bibnamefont
  {{Tsung}}}, \ and\ \bibinfo {author} {\bibfnamefont {W.~B.}\ \bibnamefont
  {{Mori}}},\ }\href {\doibase 10.1088/0741-3335/50/12/124034} {\bibfield
  {journal} {\bibinfo  {journal} {Plasma Phys. Contr. Fusion}\ }\textbf
  {\bibinfo {volume} {50}},\ \bibinfo {eid} {124034} (\bibinfo {year}
  {2008})}\BibitemShut {NoStop}%
\bibitem [{\citenamefont {Nilson}\ \emph {et~al.}(2006)\citenamefont {Nilson},
  \citenamefont {Willingale}, \citenamefont {Kaluza}, \citenamefont
  {Kamperidis}, \citenamefont {Minardi}, \citenamefont {Wei}, \citenamefont
  {Fernandes}, \citenamefont {Notley}, \citenamefont {Bandyopadhyay},
  \citenamefont {Sherlock}, \citenamefont {Kingham}, \citenamefont {Tatarakis},
  \citenamefont {Najmudin}, \citenamefont {Rozmus} \emph {et~al.}}]{Nilson06}%
  \BibitemOpen
  \bibfield  {author} {\bibinfo {author} {\bibfnamefont {P.~M.}\ \bibnamefont
  {Nilson}}, \bibinfo {author} {\bibfnamefont {L.}~\bibnamefont {Willingale}},
  \bibinfo {author} {\bibfnamefont {M.~C.}\ \bibnamefont {Kaluza}}, \bibinfo
  {author} {\bibfnamefont {C.}~\bibnamefont {Kamperidis}}, \bibinfo {author}
  {\bibfnamefont {S.}~\bibnamefont {Minardi}}, \bibinfo {author} {\bibfnamefont
  {M.~S.}\ \bibnamefont {Wei}}, \bibinfo {author} {\bibfnamefont
  {P.}~\bibnamefont {Fernandes}}, \bibinfo {author} {\bibfnamefont
  {M.}~\bibnamefont {Notley}}, \bibinfo {author} {\bibfnamefont
  {S.}~\bibnamefont {Bandyopadhyay}}, \bibinfo {author} {\bibfnamefont
  {M.}~\bibnamefont {Sherlock}}, \bibinfo {author} {\bibfnamefont {R.~J.}\
  \bibnamefont {Kingham}}, \bibinfo {author} {\bibfnamefont {M.}~\bibnamefont
  {Tatarakis}}, \bibinfo {author} {\bibfnamefont {Z.}~\bibnamefont {Najmudin}},
  \bibinfo {author} {\bibfnamefont {W.}~\bibnamefont {Rozmus}},  \emph
  {et~al.},\ }\href@noop {} {\bibfield  {journal} {\bibinfo  {journal} {Phys.
  Rev. Lett.}\ }\textbf {\bibinfo {volume} {97}},\ \bibinfo {pages} {255001}
  (\bibinfo {year} {2006})}\BibitemShut {NoStop}%
\bibitem [{\citenamefont {Li}\ \emph {et~al.}(2007)\citenamefont {Li},
  \citenamefont {Seguin}, \citenamefont {Frenje}, \citenamefont {Rygg},
  \citenamefont {Petrasso}, \citenamefont {Town}, \citenamefont {Landen},
  \citenamefont {Knauer},\ and\ \citenamefont {Smalyuk}}]{Li07}%
  \BibitemOpen
  \bibfield  {author} {\bibinfo {author} {\bibfnamefont {C.~K.}\ \bibnamefont
  {Li}}, \bibinfo {author} {\bibfnamefont {F.~H.}\ \bibnamefont {Seguin}},
  \bibinfo {author} {\bibfnamefont {J.~A.}\ \bibnamefont {Frenje}}, \bibinfo
  {author} {\bibfnamefont {J.~R.}\ \bibnamefont {Rygg}}, \bibinfo {author}
  {\bibfnamefont {R.~D.}\ \bibnamefont {Petrasso}}, \bibinfo {author}
  {\bibfnamefont {R.~P.~J.}\ \bibnamefont {Town}}, \bibinfo {author}
  {\bibfnamefont {O.~L.}\ \bibnamefont {Landen}}, \bibinfo {author}
  {\bibfnamefont {J.~P.}\ \bibnamefont {Knauer}}, \ and\ \bibinfo {author}
  {\bibfnamefont {V.~A.}\ \bibnamefont {Smalyuk}},\ }\href@noop {} {\bibfield
  {journal} {\bibinfo  {journal} {Phys. Rev. Lett.}\ }\textbf {\bibinfo
  {volume} {99}},\ \bibinfo {pages} {055001} (\bibinfo {year}
  {2007})}\BibitemShut {NoStop}%
\bibitem [{\citenamefont {Kugland}\ \emph {et~al.}(2012)\citenamefont
  {Kugland}, \citenamefont {Ryutov}, \citenamefont {Chang}, \citenamefont
  {Drake}, \citenamefont {Fiksel}, \citenamefont {Froula}, \citenamefont
  {Glenzer}, \citenamefont {Gregori}, \citenamefont {Grosskopf}, \citenamefont
  {Koenig}, \citenamefont {Kuramitsu}, \citenamefont {Kuranz}, \citenamefont
  {Levy}, \citenamefont {Liang} \emph {et~al.}}]{Kugland12}%
  \BibitemOpen
  \bibfield  {author} {\bibinfo {author} {\bibfnamefont {N.~L.}\ \bibnamefont
  {Kugland}}, \bibinfo {author} {\bibfnamefont {D.~D.}\ \bibnamefont {Ryutov}},
  \bibinfo {author} {\bibfnamefont {P.-Y.}\ \bibnamefont {Chang}}, \bibinfo
  {author} {\bibfnamefont {R.~P.}\ \bibnamefont {Drake}}, \bibinfo {author}
  {\bibfnamefont {G.}~\bibnamefont {Fiksel}}, \bibinfo {author} {\bibfnamefont
  {D.~H.}\ \bibnamefont {Froula}}, \bibinfo {author} {\bibfnamefont {S.~H.}\
  \bibnamefont {Glenzer}}, \bibinfo {author} {\bibfnamefont {G.}~\bibnamefont
  {Gregori}}, \bibinfo {author} {\bibfnamefont {M.}~\bibnamefont {Grosskopf}},
  \bibinfo {author} {\bibfnamefont {M.}~\bibnamefont {Koenig}}, \bibinfo
  {author} {\bibfnamefont {Y.}~\bibnamefont {Kuramitsu}}, \bibinfo {author}
  {\bibfnamefont {C.}~\bibnamefont {Kuranz}}, \bibinfo {author} {\bibfnamefont
  {M.~C.}\ \bibnamefont {Levy}}, \bibinfo {author} {\bibfnamefont
  {E.}~\bibnamefont {Liang}},  \emph {et~al.},\ }\href@noop {} {\bibfield
  {journal} {\bibinfo  {journal} {Nat. Phys.}\ }\textbf {\bibinfo {volume} {8}},\
  \bibinfo {pages} {809} (\bibinfo {year} {2012})}\BibitemShut {NoStop}%
\bibitem [{\citenamefont {{Davidson}}\ \emph {et~al.}(1972)\citenamefont
  {{Davidson}}, \citenamefont {{Hammer}}, \citenamefont {{Haber}},\ and\
  \citenamefont {{Wagner}}}]{Davidson72}%
  \BibitemOpen
  \bibfield  {author} {\bibinfo {author} {\bibfnamefont {R.~C.}\ \bibnamefont
  {{Davidson}}}, \bibinfo {author} {\bibfnamefont {D.~A.}\ \bibnamefont
  {{Hammer}}}, \bibinfo {author} {\bibfnamefont {I.}~\bibnamefont {{Haber}}}, \
  and\ \bibinfo {author} {\bibfnamefont {C.~E.}\ \bibnamefont {{Wagner}}},\
  }\href {\doibase 10.1063/1.1693910} {\bibfield  {journal} {\bibinfo
  {journal} {Phys. Fluids}\ }\textbf {\bibinfo {volume} {15}},\ \bibinfo
  {pages} {317} (\bibinfo {year} {1972})}\BibitemShut {NoStop}%
\bibitem [{\citenamefont {Califano}\ \emph {et~al.}(1998)\citenamefont
  {Califano}, \citenamefont {Pegoraro}, \citenamefont {Bulanov},\ and\
  \citenamefont {Mangeney}}]{Califano98}%
  \BibitemOpen
  \bibfield  {author} {\bibinfo {author} {\bibfnamefont {F.}~\bibnamefont
  {Califano}}, \bibinfo {author} {\bibfnamefont {F.}~\bibnamefont {Pegoraro}},
  \bibinfo {author} {\bibfnamefont {S.~V.}\ \bibnamefont {Bulanov}}, \ and\
  \bibinfo {author} {\bibfnamefont {A.}~\bibnamefont {Mangeney}},\ }\href@noop
  {} {\bibfield  {journal} {\bibinfo  {journal} {Phys. Rev. E}\ }\textbf
  {\bibinfo {volume} {57}},\ \bibinfo {pages} {7048} (\bibinfo {year}
  {1998})}\BibitemShut {NoStop}%
\bibitem [{\citenamefont {Silva}\ \emph {et~al.}(2003)\citenamefont {Silva},
  \citenamefont {Fonseca}, \citenamefont {Tonge}, \citenamefont {Dawson},
  \citenamefont {Mori},\ and\ \citenamefont {Medvedev}}]{Silva03}%
  \BibitemOpen
  \bibfield  {author} {\bibinfo {author} {\bibfnamefont {L.~O.}\ \bibnamefont
  {Silva}}, \bibinfo {author} {\bibfnamefont {R.~A.}\ \bibnamefont {Fonseca}},
  \bibinfo {author} {\bibfnamefont {J.~W.}\ \bibnamefont {Tonge}}, \bibinfo
  {author} {\bibfnamefont {J.~M.}\ \bibnamefont {Dawson}}, \bibinfo {author}
  {\bibfnamefont {W.~B.}\ \bibnamefont {Mori}}, \ and\ \bibinfo {author}
  {\bibfnamefont {M.~V.}\ \bibnamefont {Medvedev}},\ }\href
  {http://stacks.iop.org/1538-4357/596/i=1/a=L121} {\bibfield  {journal}
  {\bibinfo  {journal} {Astrophys. J. Lett.}\ }\textbf {\bibinfo {volume}
  {596}},\ \bibinfo {pages} {L121} (\bibinfo {year} {2003})}\BibitemShut
  {NoStop}%
\bibitem [{\citenamefont {{Schekochihin}}\ \emph {et~al.}(2009)\citenamefont
  {{Schekochihin}}, \citenamefont {{Cowley}}, \citenamefont {{Dorland}},
  \citenamefont {{Hammett}}, \citenamefont {{Howes}}, \citenamefont
  {{Quataert}},\ and\ \citenamefont {{Tatsuno}}}]{Schekochihin09}%
  \BibitemOpen
  \bibfield  {author} {\bibinfo {author} {\bibfnamefont {A.~A.}\ \bibnamefont
  {{Schekochihin}}}, \bibinfo {author} {\bibfnamefont {S.~C.}\ \bibnamefont
  {{Cowley}}}, \bibinfo {author} {\bibfnamefont {W.}~\bibnamefont {{Dorland}}},
  \bibinfo {author} {\bibfnamefont {G.~W.}\ \bibnamefont {{Hammett}}}, \bibinfo
  {author} {\bibfnamefont {G.~G.}\ \bibnamefont {{Howes}}}, \bibinfo {author}
  {\bibfnamefont {E.}~\bibnamefont {{Quataert}}}, \ and\ \bibinfo {author}
  {\bibfnamefont {T.}~\bibnamefont {{Tatsuno}}},\ }\href {\doibase
  10.1088/0067-0049/182/1/310} {\bibfield  {journal} {\bibinfo  {journal}
  {Astrophys. J. Suppl. Ser.}\ }\textbf {\bibinfo {volume} {182}},\ \bibinfo
  {pages} {310} (\bibinfo {year} {2009})}\BibitemShut {NoStop}%
\bibitem [{\citenamefont {Mondal}\ \emph {et~al.}(2012)\citenamefont {Mondal},
  \citenamefont {Narayanan}, \citenamefont {Ding}, \citenamefont {Lad},
  \citenamefont {Hao}, \citenamefont {Ahmad}, \citenamefont {Wang},
  \citenamefont {Sheng}, \citenamefont {Sengupta}, \citenamefont {Kaw},
  \citenamefont {Das},\ and\ \citenamefont {Kumar}}]{Mondal12}%
  \BibitemOpen
  \bibfield  {author} {\bibinfo {author} {\bibfnamefont {S.}~\bibnamefont
  {Mondal}}, \bibinfo {author} {\bibfnamefont {V.}~\bibnamefont {Narayanan}},
  \bibinfo {author} {\bibfnamefont {W.~J.}\ \bibnamefont {Ding}}, \bibinfo
  {author} {\bibfnamefont {A.~D.}\ \bibnamefont {Lad}}, \bibinfo {author}
  {\bibfnamefont {B.}~\bibnamefont {Hao}}, \bibinfo {author} {\bibfnamefont
  {S.}~\bibnamefont {Ahmad}}, \bibinfo {author} {\bibfnamefont {W.~M.}\
  \bibnamefont {Wang}}, \bibinfo {author} {\bibfnamefont {Z.~M.}\ \bibnamefont
  {Sheng}}, \bibinfo {author} {\bibfnamefont {S.}~\bibnamefont {Sengupta}},
  \bibinfo {author} {\bibfnamefont {P.}~\bibnamefont {Kaw}}, \bibinfo {author}
  {\bibfnamefont {A.}~\bibnamefont {Das}}, \ and\ \bibinfo {author}
  {\bibfnamefont {G.~R.}\ \bibnamefont {Kumar}},\ }\href {\doibase
  10.1073/pnas.1200753109} {\bibfield  {journal}
  {\bibinfo  {journal} {Proc. Nat Acad. Scai. U.S.A.}\ }\textbf {\bibinfo {volume} {109}},\ \bibinfo
  {pages} {8011} (\bibinfo {year} {2012})}\BibitemShut {NoStop}%
\bibitem [{\citenamefont {Gao}\ \emph {et~al.}(2013)\citenamefont {Gao},
  \citenamefont {Nilson}, \citenamefont {Igumenschev}, \citenamefont {Fiksel},
  \citenamefont {Yan}, \citenamefont {Davies}, \citenamefont {Martinez},
  \citenamefont {Smalyuk}, \citenamefont {Haines}, \citenamefont {Blackman},
  \citenamefont {Froula}, \citenamefont {Betti},\ and\ \citenamefont
  {Meyerhofer}}]{Gao13}%
  \BibitemOpen
  \bibfield  {author} {\bibinfo {author} {\bibfnamefont {L.}~\bibnamefont
  {Gao}}, \bibinfo {author} {\bibfnamefont {P.~M.}\ \bibnamefont {Nilson}},
  \bibinfo {author} {\bibfnamefont {I.~V.}\ \bibnamefont {Igumenschev}},
  \bibinfo {author} {\bibfnamefont {G.}~\bibnamefont {Fiksel}}, \bibinfo
  {author} {\bibfnamefont {R.}~\bibnamefont {Yan}}, \bibinfo {author}
  {\bibfnamefont {J.~R.}\ \bibnamefont {Davies}}, \bibinfo {author}
  {\bibfnamefont {D.}~\bibnamefont {Martinez}}, \bibinfo {author}
  {\bibfnamefont {V.}~\bibnamefont {Smalyuk}}, \bibinfo {author} {\bibfnamefont
  {M.~G.}\ \bibnamefont {Haines}}, \bibinfo {author} {\bibfnamefont {E.~G.}\
  \bibnamefont {Blackman}}, \bibinfo {author} {\bibfnamefont {D.~H.}\
  \bibnamefont {Froula}}, \bibinfo {author} {\bibfnamefont {R.}~\bibnamefont
  {Betti}}, \ and\ \bibinfo {author} {\bibfnamefont {D.~D.}\ \bibnamefont
  {Meyerhofer}},\ }\href {\doibase 10.1103/PhysRevLett.110.185003} {\bibfield
  {journal} {\bibinfo  {journal} {Phys. Rev. Lett.}\ }\textbf {\bibinfo
  {volume} {110}},\ \bibinfo {pages} {185003} (\bibinfo {year}
  {2013})}\BibitemShut {NoStop}%
\end{thebibliography}
\end{document}